\input harvmac

\input epsf

\overfullrule=0pt

\lref\vafainduc{M.\ Bershadsky, V.\ Sadov and C.\ Vafa,
{\it D-branes and Topological Field Theories},
Nucl.\ Phys.\ {\bf B463} (1996) 420-434, hep-th/9511222.}
\lref\ggs{D.\ Garfinkle, S.\ B.\ Giddings and A.\ Strominger, 
{\it Entropy in Black Hole Pair Production}, Phys.\ Rev.\
{\bf D 49} (1994) 958-965; gr-qc/9306023.}
\lref\sat{Y.\ Satoh, {\it BTZ Black Holes and the Near-Horizon Geometry of
Higher-Dimensional Black Holes}, PUPT-1816, hep-th/9810135.}
\lref\jmfd{J.\ Michelson, {\it Scattering of Four-Dimensional Black Holes},
Phys.\ Rev.\ {\bf D 57} (1998) 1092-1097; hep-th/9708091.}
\lref\dggh{F.\ Dowker, J.\ Gauntlett, G.\ Gibbons and G.\ Horowitz, 
{\it Nucleation of $p$-branes and Fundamental Strings}, Phys.\
Rev.\ {\bf D 53} (1996) 7115-7128, hep-th/9512154.}
\lref\fe{R.\ Ferrell and D.\ Eardley, {\it Slow-Motion Scattering and
Coalescence of Maximally Charged Black Holes}, Phys.\ Rev.\ Lett.\ {\bf 59}
(1987) 1617-1620.}
\lref\dkjm{D.\ M.\ Kaplan and J.\ Michelson, {\it Scattering of Several
Multiply Charged Extremal $D=5$ Black Holes}, Phys.\ Lett.\ {\bf B410} (1997)
125-130; hep-th/9707021.}
\lref\asad{A.\ Strominger, {\it $AdS_2$ Quantum Gravity and String Theory},
hep-th/9809027.}
\lref\astasi{A.\ Strominger, {\it Baby Universes}, {\bf in} TASI '88:
Particles, Strings and
Supernovae (ed.\ A.\ Jevicki and C.-I.\ Tan) World Scientific, 1989.}
\lref\ascv{A.\ Strominger and C.\ Vafa, {\it Microscopic Origin of the
Bekenstein-Hawking Entropy},
 { Phys.\ Lett.\ } {\bf B379} (1996) 99-104, hep-th/9601029.}
\lref\larsen{M.\ Cveti\v{c} and F.\ Larsen, {\it Microstates of Four
Dimensional Black Holes from Near-Horizon Geometry}, UPR-798-T,
ITP/NSF-98-065, hep-th/9805146\semi
M. Cveti\v{c} and F. Larsen, 
{\it Greybody Factors for Black Holes in Four Dimensions}, 
Phys. Rev. {\bf D 57} (1998) 6297-6310, hep-th/9712118\semi
M. Cveti\v{c} and F. Larsen, 
{\it Greybody Factors for Rotating Black Holes in Four Dimensions},
Nucl. Phys. {\bf B506} (1997) 107-120,  hep-th/9706071. 
}
\lref\jm{J.\ Maldacena, 
{\it The Large $N$ Limit of Superconformal Field
Theories and Supergravity},
Adv. Theor. Math. Phys. {\bf 2} (1998) 231-252,
hep-th/9711200.}
\lref\jmasu{J.\ Maldacena and A.\ Strominger,
{\it Universal Low-Energy Dynamics for Rotating Black Holes},
Phys.\ Rev.\  {\bf D 56} (1997) 4975-4983, hep-th/9702015.}
\lref\dpop{M.\ J.\ Duff, H.\ L\"{u} and C.\ Pope, 
{\it $AdS_3 \times S^3$ (Un)twisted and Squashed, and an $O(2,2;\IZ)$
Multiplet of Dyonic Strings}, CTP TAMU-28/98, LPTENS-98/30,
SISSA~Ref.~79/98/EP, hep-th/9807173.} 
\lref\msfour{J.\ Maldacena and A.\ Strominger, 
{\it Statistical Entropy of Four-Dimensional Black Holes},
Phys.\ Rev.\ Lett.\ {\bf 77}
(1996) 428-429, hep-th/9603060.}
\lref\psstw{J.\ Preskill, P.\ Schwarz, A.\ Shapere, S.\ Trivedi and F.\
Wilczek, {\it Limitations on the Statistical Description of Black Holes},
Mod.\ Phys.\ Lett.\ {\bf A6} (1991) 2353-2362.}
\lref\sknd{H.\ J.\ Boonstra, B.\ Peeters and K.\ Skenderis, 
{\it Brane intersections, Anti-de Sitter Spacetimes and Dual Superconformal
Field Theories}, Nucl.\ Phys.\
{\bf B533} (1998) 127-162, hep-th/9803231.}
\lref\msu{
J. Maldacena and L. Susskind, {\it D-branes and Fat Black Holes},
Nucl.\ Phys.\ {\bf B475} (1996) 679-690, hep-th/9604042.}
\lref\brl{D.\ Brill, {\it Splitting of an Extremal Reissner-Nordstr\"{o}m Throat
via Quantum Tunneling}, Phys.\ Rev.\ {\bf D46} (1992) 1560-1565.}
\lref\zerotwo{
 O.\ Aharony, M.\ Berkooz, S.\ Kachru, N.\ Seiberg and  E.\
Silverstein,
{\it Matrix Description of Interacting Theories in Six Dimensions},
Adv.\ Theor.\ Math.\ Phys.\ {\bf 1} (1998) 148-157,
hep-th/9707079. } 
\lref\higgsdecoupling{ E.\ Witten,
{\it On the Conformal Field Theory of the Higgs Branch}, 
JHEP {\bf 07} (1997) 003, hep-th/9707093.}
\lref\instantons{
T.\ Banks and M.\ Green, 
{\it Nonperturbative Effects in $AdS^5\times S^5$ String Theory and
$d=4$ SUSY Yang-Mills}, 
JHEP {\bf 05} (1998) 002, 
hep-th/9804170\semi
C.\ S.\ Chu, P.\ M.\ Ho and  Y.\ Y.\ Wu,
{\it D-Instanton in $AdS_5$ and Instanton in SYM$_4$}, SISSA-58/98/FM,
SHU-TIPAC-98007,
 hep-th/9806103\semi
  I.\ Kogan, G.\ Luzon, {\it D-Instantons on the Boundary}, OUTP-98-49P,
hep-th/9806197\semi 
 V.\ Balasubramanian, P.\ Kraus, A.\ Lawrence and S.\ Trivedi,
{\it Holographic Probes of Anti-de Sitter Spacetimes},
HUTP-98/A057, CALT68-2189, Fermilab-Pub-98/240-T,
hep-th/9808017\semi
 M.\ Bianchi, M.\ Green, S.\ Kovacs and  G.\ Rossi,
{\it Instantons in Supersymmetric Yang-Mills and D-Instantons in IIB 
Superstring Theory},
 JHEP {\bf 08} (1998) 013, hep-th/9807033.} 
\lref\teitel{J.\ Brown and C.\ Teitelboim, {\it Neutralization of the
Cosmological Constant by Membrane Creation}, Nucl.\ Phys.\ {\bf B297} (1988)
787-836.}
\lref\warp{ P.\ Van Nieuwenhuizen and N.\ Warner, {\it 
New Compactifications of Ten Dimensional and Eleven Dimensional 
Supergravity on Manifolds Which Are Not Direct Products},
Comm.\ Math.\ Phys.\ {\bf 99} (1985) 141\semi
C.\ Hull and N.\ Warner, {\it 
Noncompact Gaugings From Higher Dimensions}, 
Class.\ Quant.\ Grav.\ {\bf 5} (1988) 1517.}
\lref\sofive{ M.\ G\"{u}naydin, L.\ Romans and N.\ Warner,
{\it Compact and Noncompact Gauged Supergravity Theories in Five 
Dimensions}, Nucl.\ Phys.\ {\bf B272} (1986) 598\semi
L.\ Girardello, M.\ Terini, M.\ Porrati and A.\ Zaffaroni, 
{\it Novel Local CFT and Exact Results on Perturbations
of $N=4$ Super Yang Mills From $AdS$ Dynamics}, 
CERN-TH/98-323, IFUM-633-FT, IMPERIAL/TP/98-99/4, NYU-TH/98/10/03,
hep-th/9810126\semi
J.\ Distler and F.\ Zamora,
{\it Non-supersymmetric Conformal Field Theories From Stable 
Anti-de Sitter Spaces}, UTTG-14-98, hep-th/9810206.}

\def\st{$S^2$}
\def\vx{{\vec{x}}}
\def\vu{{\vec{U}}}
\def\adst{$AdS_2\times S^2$}

\def\ads{$AdS_3$}
\def\ad{$AdS_2$}

\def\slr{{SL(2,\IR)}}

\def\p{\partial}

\def\rb{Robinson-Bertotti}
\def\IR{\relax{\rm I\kern-.18em R}}
\def\IZ{\relax\ifmmode\hbox{Z\kern-.4em Z}\else{Z\kern-.4em Z}\fi}
\Title{\vbox{\baselineskip12pt\hbox{HUTP-98/A088}
\hbox{UCSBTH-98-8}\hbox{hep-th/9812073}}}{Anti-de Sitter Fragmentation}

\centerline{Juan Maldacena$^1$, Jeremy Michelson$^{2,1}$ and Andrew  Strominger$^1$}
\bigskip\centerline{$^1$Department of Physics}
\centerline{Harvard University}
\centerline{Cambridge, MA 02138} 

\bigskip\centerline{$^2$Department of Physics}
\centerline{University of California}
\centerline{Santa Barbara, CA  93106}

\vskip .3in
\centerline{\bf Abstract}
Low-energy, near-horizon scaling limits of black holes 
which lead to 
string theory on \adst\ are described. Unlike the higher-dimensional 
cases, in the simplest approach
all finite-energy excitations of \adst\ are suppressed. 
Surviving zero-energy 
configurations are described. These can include tree-like 
structures in 
which the \adst\ throat branches as the horizon is approached, 
as well as disconnected \adst\ universes.  In principle, the black
hole entropy counts the quantum ground states on the moduli 
space of such configurations. In a nonsupersymmetric context 
$AdS_D$ for general $D$ can be unstable against 
instanton-mediated fragmentation 
into disconnected universes. Several examples are given. 
\smallskip
\Date{}
\newsec{Introduction}
   By now a beautiful and coherent story has been developed for the
$AdS_D/CFT_{D-1}$ duality \jm\ for several values of $D$. A notable 
exception is the enigmatic case $D=2$. This case is 
perhaps the most interesting from the point of view of black hole physics
because it is the very-near-horizon geometry of all known cases of
supersymmetric black holes with non-zero entropy.\foot{In some cases one encounters quotients 
of \ads, but these also reduce to \ad\ at sufficiently low energies.\asad}  
One immediately puzzling feature is
the fact that \ad\ has two disconnected boundaries. From this 
alone it is evident that the $D=2$ case must involve qualitatively new
features.  Some preliminary progress on this
case was reported in \refs{\jmasu,\larsen,\sknd,\asad,\sat} .

In this paper we shall continue exploration of the $AdS_2/CFT_1$ 
duality. In section 2 we analyze several inequivalent 
approaches to the near-horizon limit. We shall see that 
it is $not$ possible to keep the charge, energy and temperature fixed 
in the usual manner 
while taking the Planck mass $M_p \to \infty$. In the most straightforward 
near-horizon limit the excitation energy of \ad\ is forced to zero. 
The resulting theory describes only the (many) extremal 
black hole ground states. 

\ad\ is not the only zero-energy configuration which survives the
$M_p \to \infty$ limit. In addition one can have geometries which are 
asymptotic to \ad\ at large radius but branch (in a tree-like structure) 
into smaller \ad\ regions as 
one moves toward the horizon. In section 3 we describe these 
configurations and their low energy dynamics. We further discuss 
their description in the dual CFT.

In section 4 we consider general $AdS_D$ spaces and discuss
the possibility of brane creation by the antisymmetric D-form field 
strength \refs{\teitel,\dggh}. If
$D>2$, it can only occur in non-supersymmetric cases. We present an example of a non-supersymmetric $AdS_3\times S^3 \times
K3$ compactification where this seems to be the dominant decay mode. 
Finally we discuss 
topology changing instantons in the supersymmetric $AdS_2$ case. 
These instantons, found by Brill \brl, describe tunneling between 
several $AdS_2$ spaces.

\newsec{\ad\ as a Low-Energy Limit}

The oldest and simplest example of \ad\ arising as a near-horizon geometry
is in the context of the Reissner-Nordstr\"{o}m solution of four-dimensional 
Einstein-Maxwell gravity. The full magnetically-charged solution is 
\eqn\reno{\eqalign{ds^2&=- {(r-r^+)(r-r^-) \over r^2}dt^2 +{r^2 \over (r-r^+)(r-r^-)}dr^2
+r^2 d\Omega_2^2,\cr  F&=Q\epsilon_2, \cr  r_\pm&=QL_p+EL_p^2 \pm\sqrt{2QEL_p^3 +E^2L_p^4}.}}
In this expression $L_p$ is the Planck length ($L_p^2 = G_N$), 
$\epsilon_2$ is the volume element on the 
unit $S^2$ and 
\eqn\dsf{E=M-{Q \over L_p}}
is the excitation energy above extremality. String theoretic examples 
generically involve more fields and several charges. We will 
mainly consider the Reissner-Nordstr\"{o}m  example because it is 
simpler and has qualitatively similar behavior. 

An important feature of these black holes \psstw\ is that 
the semiclassical analysis of their thermodynamic behavior breaks 
down very near extremality.  
This follows from the formulae for the  entropy and Hawking temperature 
\eqn\etl{\eqalign{S_{BH}&={ \pi r_+^2 \over L_p^2},\cr 
T_H&= {r_+-r_- \over 4 \pi r_+^2}.}}
Near extremality the energy-temperature relation is 
\eqn\opy{E\sim 2 \pi^2 Q^3 T_H^2L_p.}
The energy of a typical quantum of Hawking emission is of order $T_H$.
When this energy is of order of or greater than the total 
available energy $E$ above extremality, 
the semiclassical analysis must break down. This occurs at an 
excitation energy of order 
\eqn\egp{E_{gap} \equiv {1 \over Q^3 L_p}.}
In string theory examples the nature of this breakdown is 
well-understood \msu : the black hole has a mass gap and \egp\ 
is the energy of its lowest-lying excitation\foot{There are however 
lower-energy 
modes describing the fragmentation of the black hole into smaller 
pieces. These modes
are discussed below.}. In the description of a 
four-dimensional black hole given in \msfour\ the gap state is the 
lowest excitation of a conformal field theory on a circle. In more general 
stringy constructions of four-dimensional black holes it is not 
always possible to 
compute the gap but the semiclassical analysis of \psstw\ indicates it will 
always be of the order \egp.

The near-horizon limit is simplest to describe for the extremal 
case in which $E=0$, $r_+=QL_p$ and $T_H=0$. One then considers 
the limit
\eqn\nhr{L_p \to 0,}
with 
\eqn\fgh{U= {r-r_+ \over L_p^2},~~~~~~~Q~~  {\rm fixed} .}
The metric then reduces to    
\eqn\dfr{{ds^2\over L_p^2} =- { U^2\over Q^2}dt^2 +{Q^2 \over U^2}dU^2
+Q^2 d\Omega_2^2.}
In null coordinates 
\eqn\rgo{u^\pm={\rm arctan} (t\pm {Q^2 \over U}),}
the metric is 
\eqn\dffr{{ds^2\over L_p^2} =-{4Q^2du^+du^- \over \sin ^2 (u^+-u^-)}
+Q^2 d\Omega_2^2.}
This is known as the \rb\ geometry on \adst . As illustrated in figure 1, 
the \adst\ region of the full Reissner-Nordstr\"{o}m geometry is a ribbon which zigzags 
its way up through the infinite chain of universes. One of the timelike 
boundaries of \ad\ ($u^+=u^-$) is just outside the black hole horizon, 
while the other ($u^+=u^-+\pi$) is just inside. 

\vskip .6cm
\centerline{
\epsfxsize=0.3\hsize\epsfbox{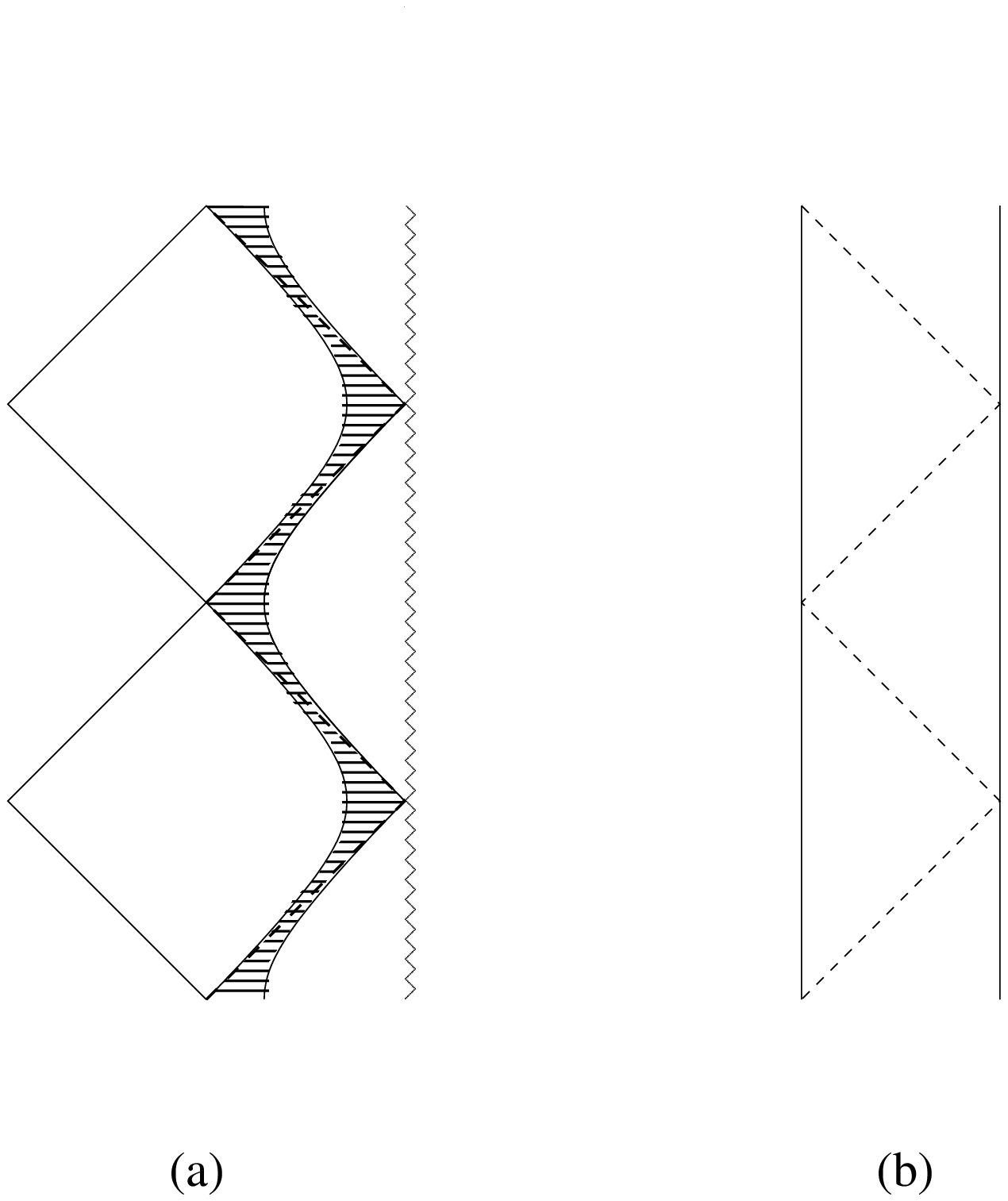}}
\bigskip
\centerline{\vbox{\noindent {\bf Fig.\ 1.} 
 (a) Penrose diagram corresponding to the extremal 
Reissner-Nordstr\"{o}m black hole. The dashed line is the black hole
horizon, and the shaded strip is the near-horizon
$AdS_2$ region. (b) Penrose diagram for $AdS_2$. The diagonal 
lines are the horizons inherited from the embedding in 
Reissner-Nordstr\"{o}m. 
}}
\vskip .7cm

Since there are two timelike boundaries, quantum gravity with fermions 
on \ad\ will have a NS and  R sector. In the supersymmetric case, 
the Hamiltonian $H_0$ which generates time translations of the full solution
\reno\ is the square of a supercharge: $H_0=Q^2$. This Hamiltonian generates accelerating 
trajectories in the global coordinates \dffr\ of the near-horizon \ad. 
The NS sector has a supercharge ($G_{1/2} +G_{-1/2}$) which squares to $H_0$, 
but the R sector does not. Therefore in the supersymmetric case the 
near-horizon limit leads to the NS sector of quantum gravity on \ad.  

One may also wish to consider  more general limits  which 
are not restricted to zero temperature and excitation energy. This problem is 
qualitatively different than its higher-dimensional cousins because of the 
explicit factors of $L_p$ appearing in \opy\ and \egp. In the following 
subsections we consider four such more general limits. 

\subsec{{\bf Limit \# 1 :}~~$L_p \to 0,~~(E, ~Q)$ Fixed}

The limit $L_p \to 0$ with $(E, ~Q)$ fixed is problematic because according to 
\opy\ the Hawking temperature is infinite and the geometry is singular. 
Hence we do not know how to make sense of this limit. Note that for higher 
($p>0$) $p$-branes with $AdS$ horizons 
the energy-temperature relation extracted from the 
near-extremal solutions  is of the form 
\eqn\etr{E \sim V_pT_H^{p+1}.}
This involves no explicit factors of $L_p$ (by dimensional analysis)
and the $L_p \to 0$ limit is nonsingular for either 
fixed temperature or energy.
In contrast the energy-temperature relation \egp\ is characteristic 
of a string-like 
rather than a point-like object. In place of the missing brane 
dimension a power of 
$L_p$ appears in \opy . 
  
\subsec{{\bf Limit \# 2 :}~~$L_p \to 0,~~(T_H, ~Q)$ Fixed}

In order to keep the geometry of the solution fixed as 
$L_p \to 0$, one should keep the temperature $T_H$, which is 
related to the periodicity of the solution, fixed. From \opy\ it 
immediately follows that the excitation energy $E$ 
vanishes in such a limit, while the gap energy \egp\ goes to infinity. 
Defining $U$ exactly as in \fgh\ one finds that the metric reduces to    
\eqn\dfrg{{ds^2\over L_p^2} =- { U(U+4\pi Q^2T_H)\over Q^2}dt^2 
+{Q^2 \over U(U+4\pi Q^2 T_H)}dU^2
+Q^2 d\Omega_2^2.}
The metric \dfrg\ takes the canonical form \dfr\ in 
the primed coordinates
\eqn\ctf{t^\prime \pm {Q^2 \over U^\prime} = \tanh \left[ \pi T_H \left(t \pm {1 \over
4 \pi T_H} \ln {U \over U+4 \pi Q^2 T_H}\right) \right]}
Hence at the classical level the geometry of the
$L_p \to 0$ limit is independent of $T_H$. 

At the quantum level there is 
a distinction. The quantum vacuum depends on the choice of 
time coordinate used to distinguish positive and negative frequency 
oscillators. The $\slr$ invariant vacuum leads to a thermal bath of
particles in the 
coordinates \dfrg.  The energy density of this thermal bath 
for the case of a 
conformal matter system with 
central charge $c$ is 
\eqn\rtj{T_{00}={c\pi T_H^2\over 6}.}

The impossibility of non-singular 
finite energy excitations of \adst\ at the classical
level can be seen directly from the classical equations. 
The classical action in two dimensions contains the gravity-dilaton 
terms 
\eqn\trm{S_2={1 \over 4}\int d^2 x\sqrt{-g}\left
\{e^{-2 \phi}[R
-F^2 + 2 (\nabla \phi)^2]+{2\over
L_p^2} \right\}+S_M.}
where $S_M$ is the matter action, $R$ is the 
two dimensional scalar curvature, $4\pi e^{-2\phi}$
is the volume of the $S^2$ and $F=dA$ is the gauge field 
strength.
The $++$ constraint equation can be written  
\eqn\cstr{-2e^{-\phi}\nabla_+\nabla_+ e^{-\phi}=T_{++}\ge 0.}
Integrating \cstr\ in conformal gauge $ds^2=e^{2\rho} du^+du^-$ with the
measure $e^{\phi -2\rho}du^+$ across \ad\ from $0$ to $\pi$ 
along the line $u^-=0$ gives
\eqn\drl{e^{-2\rho}\p_+e^{-\phi}|_{u^+=0}- 
e^{-2\rho}\p_+e^{-\phi}|_{u^+=\pi}
={1 \over 2}\int du^+e^{\phi-2\rho}T_{++}\ge 0.}
$e^{-2\rho}$ vanishes quadratically near the boundaries for \ad . 
If $T_{++}$ is nonzero, then \drl\ implies that 
$e^{-\phi}$ must diverge linearly near at least one 
of the two boundaries.\foot{The rate of divergence  
depends on the coefficients in \trm\ which vary among different examples.} Hence the geometry cannot be asymptotic to \adst\ 
when $T_{++}$ is nonzero.    

This zero-energy constraint might be
modified at the quantum level in order to account for 
the energy of the Hawking radiation. 
Indeed the quantum constraints 
contain a one-loop correction from $c$ massless matter fields 
of the form 
${c \over 12}(\p_+^2 \rho-(\p_+\rho)^2)$ which modifies the preceding analysis. 

The classical argument does not eliminate the possibility of
black-hole-like spacetimes 
which have non-singular spacelike slices, but contain spacelike singularities in the past 
and future. For such spacetimes there may not be null surfaces which cross from one boundary to the other. A similar analysis using the time-time constraint 
equations on spacelike slices may yield relevant information, 
but one must consider the 
possibility of negative $T_{00}$ from tachyons.  

Although the energy is classically constrained to vanish for nonsingular 
spacetimes, this limit is
far from trivial because it should retain all 
the ground states of the 
system. The ground state entropy is $S=\pi Q^2$ for large $Q$. 
We have not succeeded in understanding the proper description of 
all these states in this framework. One possibility, discussed below, 
is that they arise from modes corresponding to black hole 
fragmentation or the separation of 
\adst\ universes.

\subsec{{\bf Limit \# 3 :}~~$L_p \to 0,~~(E, ~T_H)$ Fixed, $Q \to \infty$ }

  In \jmasu\ it was found that Green functions in the 
near-horizon \adst\ geometry 
of the Reissner-Nordstr\"{o}m black hole agreed with those 
of a $1+1$ chiral conformal field theory. The detailed agreement 
persisted when angular momentum was added to the black hole \jmasu\ 
and also for general charges \larsen .  Since Green functions 
measure correlations of finite-energy disturbances these  
results suggest 
the existence of an $AdS_2/CFT_1$ correspondence involving 
non-zero excitation energies. This may seem to contradict the 
analysis of the previous subsection. However a somewhat different limit was
implicit in \jmasu. One can hold both $E$ and $T_H$, as well as $E_{gap}$, 
fixed as $L_p \to 0$ at the price of taking the charge $Q \sim L_p^{-1/3} 
\to \infty$. Since $Q$ is diverging this is a large $N$ limit.
Defining 
\eqn\typ{V={r-r_+ \over Q^2L_p^2},}
The metric \reno\ in the limit $L_p \to 0$ with fixed $T_H$, $E$ 
and $E_{gap}$ defined in \egp\ takes the \adst\ form 
\eqn\drfn{ {( E_{gap} L_p)^{2/3}  \over L_p^2} 
ds^2 =- { V(V+ 4 \pi  T_H)}dt^2 
+{1 \over V(V + 4 \pi  T_H)}dV^2
+d\Omega_2^2.}

The problem of small energies leading to divergent $\phi$
does not appear because $e^{-2\phi}$, which appears on the 
left hand side of \cstr , is of order $ Q^2 $
and diverges for $L_p \to 0$. The right hand side is 
kept finite of order $E$ and can be neglected in comparison. 
The back reaction of matter on $\phi$ is suppressed. 
In string theory one also finds that the massive string modes 
decouple in this limit. Hence the limit largely consists of 
the free supergravity on \adst. Nevertheless as seen in 
\jmasu\ an $AdS/CFT$ duality already has nontrivial content 
within this limit.

\subsec{{\bf Limit \# 4 :}~~$L_p ~~small,~~(E,~T_H,  ~Q)$ Fixed}

Let us consider an infrared cutoff $U/Q^2   <  \Lambda $ on \adst. 
In the dual CFT description this
should correspond to an ultraviolet   cutoff proportional to 
$\Lambda $. The cutoff theory should be capable of describing 
states with energies $E\ll { \Lambda }$. 
It follows from the discussion in section 2.2
that the addition of energy to \adst\ produces a dilaton 
which grows like $ { E  U  L_p^2/  Q^4 } $ 
for large  $U $ and small $E$.
So if $E$ is small we can choose a cutoff which satisfies 
$E \ll \Lambda \ll Q/L_p $ so that the dilaton is small
for all $ U/Q^2 < \Lambda$. 
Hence there should be a potentially useful 
approximate duality relating the cutoff theories at very low energies.

\newsec{\ad\ Trees }

The \adst\ geometry \dfr\ is not the unique classical 
charge $Q$, $E=0$  configuration which 
survives the  $L_p \to 0 $ limit  with fixed $Q$. In addition there
are classical solutions corresponding to BPS-saturated 
multi-black hole solutions with separations of order $L_p^2$
which survive as distinct objects in the limit $L_p \to 0$. These ``tree''
geometries are asymptotically \adst\ at large radius, but as one moves 
inward the geometry branches into smaller \adst\ regions. 
In a supersymmetric theory the quantum ground states are 
cohomology classes on the moduli space of such solutions. In principle 
one might understand 
the extremal black hole entropy by counting cohomology classes on this space. 
In this section we will discuss these configurations.

\subsec{Two-Black Hole Configurations}

Let us begin with the asymptotically flat solution describing two 
Reissner-Nordstr\"{o}m black holes (see fig.\ 2) 
\eqn\ssd{\eqalign{ds^2&=-V^{-2}dt^2 +V^2d\vx^2, \cr
                   *F&={1 \over L_p} dt\wedge dV^{-1},\cr
V&=1+{Q_1 L_p \over |\vx -\vx_1|}+{Q_2 L_p \over |\vx -\vx_2|.}}}
Defining 
\eqn\rfl{\eqalign{\vu &={\vx \over L_p^2},\cr
\vu_1 &={\vx_1 \over L_p^2},\cr
 \vu_2 &={\vx_2 \over L_p^2},\cr}}
and taking $L_p \to 0$, the near-horizon metric 
becomes 
\eqn\ssdt{\eqalign{{ds^2 \over L_p^2}&=-V^{-2}dt^2 +V^2d\vx^2, \cr
                   *F&=dt\wedge dV^{-1},\cr
V&={Q_1 \over |\vu -\vu_1|}+{Q_2 \over |\vu -\vu_2|}.}}

\vskip .6cm
\centerline{
\epsfxsize=0.4\hsize\epsfbox{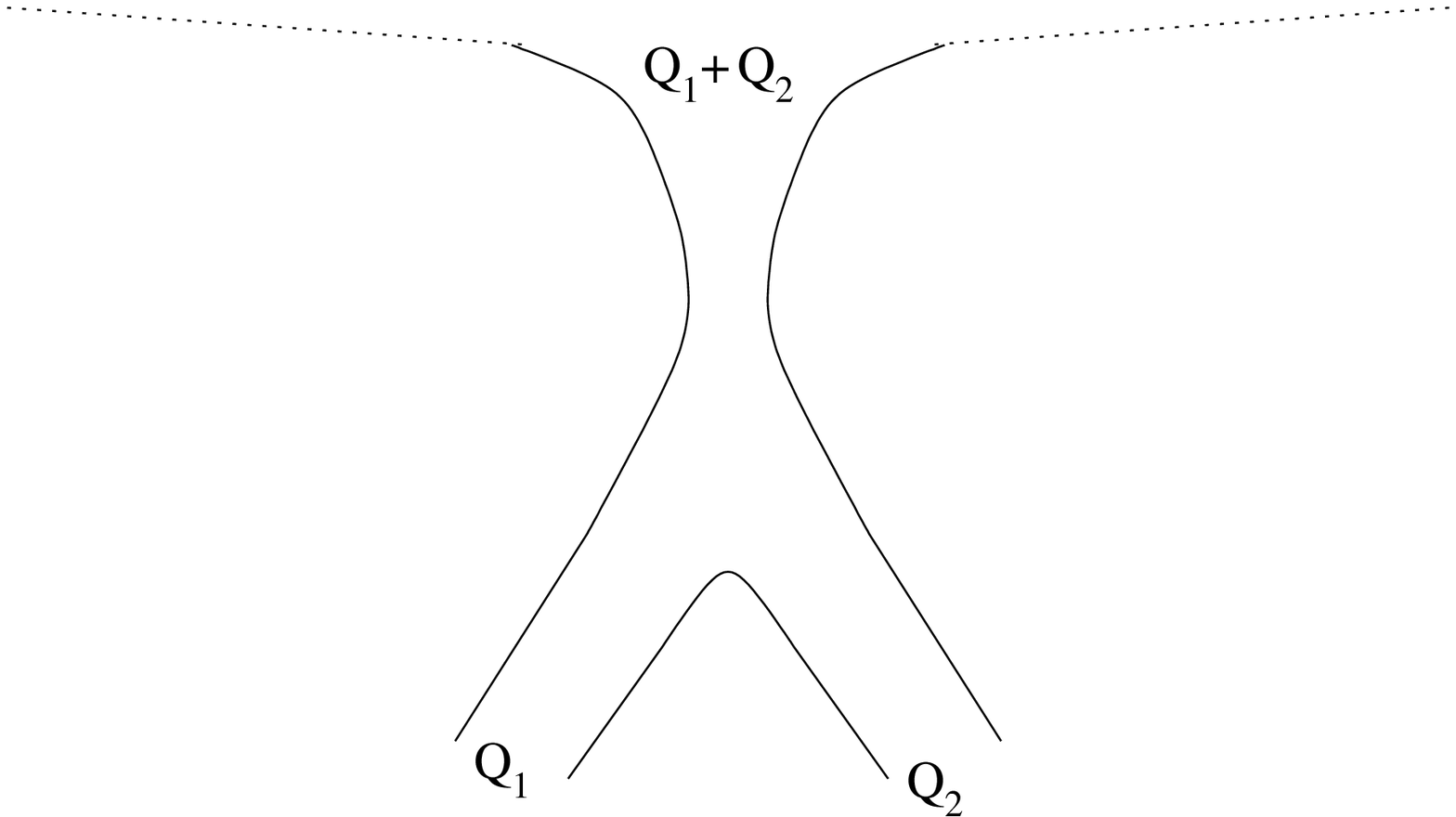}}  
\bigskip
\centerline{\vbox{\noindent {\bf Fig.\ 2.} 
A spatial cross-section the metric \ssd . There is an  asymptotically  
Minkowskian region and a single charge $Q_1+Q_2$ throat region 
which divides into two throats of charges $Q_1$ and $Q_2$. 
In the $M_p \to \infty$ limit \ssdt\ the throat 
becomes infinitely long and the Minkowski region decouples. 
The splitting of the throat into two pieces survives this limit. 
}}
\vskip .7cm

The difference $\vu_{12}=\vu_1-\vu_2$ is a collective coordinate 
for the solution. The effective action for small variations of 
this coordinate is easily obtained by scaling the known result~\fe\
\eqn\erf{S_{12}=\half {(Q_1^3 Q_2 + Q_1 Q_2^3)}\int dt {(\p_t\vu_{12})^2 \over
|\vu_{12}|^3},} 
and is finite for $L_p \to 0$. 
This geometry is locally flat with a conical singularity. It has the 
perhaps counterintuitive feature that the volume of the moduli space for 
widely separated black holes is very small, while that of nearly coincident
black holes is divergent. The region 
corresponding to widely separated
black holes is the point $|\vu_{12}|\to \infty$ at the apex of the cone.
Nearby black holes occupy the locally asymptotically flat region 
$|\vu_{12}|\to 0$. It is possible that black hole entropy can be understood as the zero-energy 
states at the boundary of this moduli space.

Naively this divergent volume leads to 
an infinite number of arbitrarily low-lying excitations of 
near-coincident black holes. A similar divergence appears 
in counting the low-energy modes of a free scalar field in the 
vicinity of a black hole 
horizon because of arbitrarily large near-horizon redshifts. 
Presumably higher order corrections regulate the divergence in both 
cases, but we do not understand how this comes about.

\subsec{The Zerobrane Limit}

It is instructive to consider the case $Q_1{\ll}Q_2$. One can then view 
the charge $Q_1$ black 
hole as a charged BPS zerobrane in \adst\ with a constant electric field. 
The coupled action for 
such a zerobrane in \ad\ is 
\eqn\strt{\eqalign{S_2=&{1 \over 4}\int d^2 x\sqrt{-g}\left
\{e^{-2 \phi}[R
-F^2 + 2 (\nabla \phi)^2]+{2\over
L_p^2} \right\}\cr& + \half \oint dx \sqrt{h} e^{-2 \phi} 
K 
+ {Q_1 \over L_p} \int A -  {Q_1 \over L_p} \int ds.}}
The scalar $\phi$ measures the size of the $S^2$ and the gauge field 
strength $F=dA$ is such that $e^{-2 \phi} F$ is of order $Q_2$.  The metric
on the boundary of \ad\ is $h$ and 
$K$ is the extrinsic curvature of the boundary of $AdS$.
The last two terms are the worldline action 
of the zerobrane.

Taking \ssdt\ and averaging $\vec x_1$ over the two sphere we get
\eqn\mjo{\eqalign{{ds^2 \over L_p^2} &= -{U^2 \over Q_2^2} {1 \over h(U)^2}
dt^2 + {Q_2^2 \over U^2} h(U)^2 dU^2, \cr
e^{-\phi} &= Q_2 h(U), \cr
{A \over L_p} &= {U \over Q_2} {1 \over h(U)} dt,}}
where
\eqn\defh{h(U) = 1+{Q_1 \over Q_2} \left(\Theta(U-a)+{U \over a}
\Theta(a-U) \right),}
where $\Theta(U)$ is the Heaviside step function.
This represents a spherical  distribution of  zerobranes of total 
 charge $Q_1$ hovering a 
fixed distance
 from the horizon. 
The zerobrane worldline is at $U=a$, where $a$ is the collective
coordinate. 

The solution \mjo\
covers only the region outside the horizon. 
The analytic extension of this metric
is provided by the Eddington-Finkelstein coordinates%
\foot{The finite coordinate transformation is 
\eqn\finiteef{\bar{t}-\bar{t}_0 = t - U - {Q_2^2 \over
U} h(U)^2 - 2 \left({Q_1^2 \over a^2} (a-U) - {Q_1 Q_2 \over a} \ln {U
\over a} \right) \Theta(a-U).}}
\eqn\defef{d\bar{t} = dt - \left(1- {Q_2^2 \over U^2} h(U)^2\right) dU.}
In these coordinates the solution is
\eqn\gld{\eqalign{{ds^2 \over L_p^2} &= - {U^2 \over Q_2^2}
{1 \over h(U)^2} d\bar{t}^2 + 2 \left(1-{U^2 \over Q_2^2}{1 \over h(U)^2}\right)
d\bar{t} dU + \left(2-{U^2 \over Q_2^2}{1 \over h(U)^2}\right) dU^2,\cr
e^{-\phi} &= Q_2 {h(U)} \cr
{A \over L_p} &= {U \over Q_2}{1 \over h(U)} 
d\bar{t}
.}} 
Note that this is now regular at $U=0$,
so we have extended the solution to
$U<0$.  However, at $U=-Q_2 a/Q_1$, there is a singularity in the metric.
This is an essential singularity of the solution; not only does the scalar
curvature diverge at this point, but $e^{-\phi}$, which is
the size of the internal $S^2$, is degenerating there (from~\defh).
This is an important contradistinction with pure \adst.

Empty \adst\ has an inner and an outer boundary, both of which are 
nonsingular.  We have just seen that \gld\ has the feature that the inner
boundary is singular in 
our semiclassical description (see Figure 3). 
This suggests that the dual one-dimensional 
conformal field theory lives on the outer boundary alone, 
while some appropriate boundary conditions must be found to define the 
dynamics of the inner boundary. 

\vskip .6cm
\centerline{
\epsfxsize=0.1\hsize\epsfbox{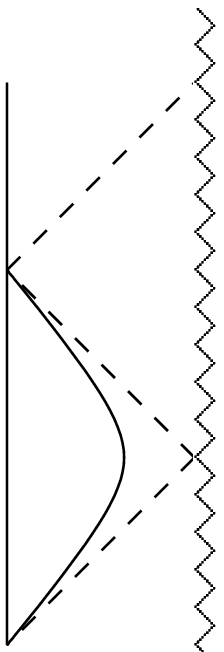}}
\bigskip
\centerline{\vbox{\noindent {\bf Fig.\ 3.} 
The Penrose diagram for the solution~\gld.  The curved path is the
zero-brane worldline $U=a$, and the dashed line is the horizon $U=0$.
}}
\vskip .7cm

Another way to understand the appearance of a singularity in \gld\ is to 
notice that for $U<a$ the harmonic function appearing in the
metric \mjo\ is of the form $ h(U)/U  \sim 1/U + const $ so it is of
the same form as that of the extremal Reissner-Nordstr\"{o}m black hole before 
the near-horizon limit is taken. 

\subsec{Charged Geodesics}

In the limit in which the back reaction of the zerobrane on the \adst\ geometry 
is neglected, it obeys a charged geodesic equation. This equation can be 
solved in general. We  consider first 
the Euclidean case which is relevant for 
the discussion of instantons in section 4. 

It is easiest to calculate the zerobrane trajectories in 
Poincare coordinates
\eqn\pnc{ds^2={dt^2+dy^2 \over y^2}.} 
where the worldline action in \strt\ is 
\eqn\euclpoin{
S= m \int dt {{ \sqrt{ 1 + \left({ d y \over dt }\right)^2 } 
-1 }\over y }.
}
The classical solutions to \euclpoin\ are 
circles of arbitrary radius which are tangent to the boundary
\eqn\geopoin{
(t-t_0)^2 + (y-a)^2 = a^2
}
as shown in fig.\ 4a.  There is an additional solution
\eqn\zeroenpoing{y = y_0.}
These solutions are transformed into one another by the action of $\slr$.
In the strip coordinates
\eqn\strp{ds^2={d\tau^2+d\sigma^2 \over \sin^2\sigma},}
the action becomes
\eqn\euclstrip{
S = m \int d\tau 
{{ \sqrt{ 1 + \left({ d \sigma  \over d \tau }\right)^2 }
 - \cos\sigma }\over \sin \sigma }
}
In this case the solutions are generically 
closed curves tangent to the boundary.  For $t_0>0$, these are tangent
to $\sigma =0$, and can be written
\eqn\geoeucl{
 \cosh{(\tau-\tau_0)} = {\cos(\sigma_m/2 -\sigma) \over \cos\sigma_m/2 }
}
as shown in fig.\ 4b.  The $t_0<0$ solutions, tangent to $\sigma=\pi$, are
mirror images of fig.\ 4b,
including orientation.
The additional solution is (see fig.\ 4c)
\eqn\zeroenstrip{
e^{-(\tau -\tau_0)} = \sin\sigma.
}
This trajectory has zero Euclidean energy and is relevant 
to vacuum tunneling.  The time-reverse of~\zeroenstrip\ is also a geodesic,
corresponding to~\geopoin\ with $t_0=0$.

Lorentzian trajectories can be obtained by Wick rotation. 
In Poincare coordinates they are (see fig.\ 4d)
\eqn\lorpoinc{ 
-(t-t_0)^2 + (y+a)^2 = a^2
}
Similarly in the strip coordinates we get (see fig.\ 4e)
\eqn\lorstrip{
\cos(\tau-\tau_0) = { \sin(\sigma_m/2-\sigma )\over \sin(\sigma_m/2)}
}
and its mirror image.  Note that the particle gets to the boundary in
finite global time. 

\vskip .6cm
\centerline{
\epsfxsize=.8\hsize\epsfbox{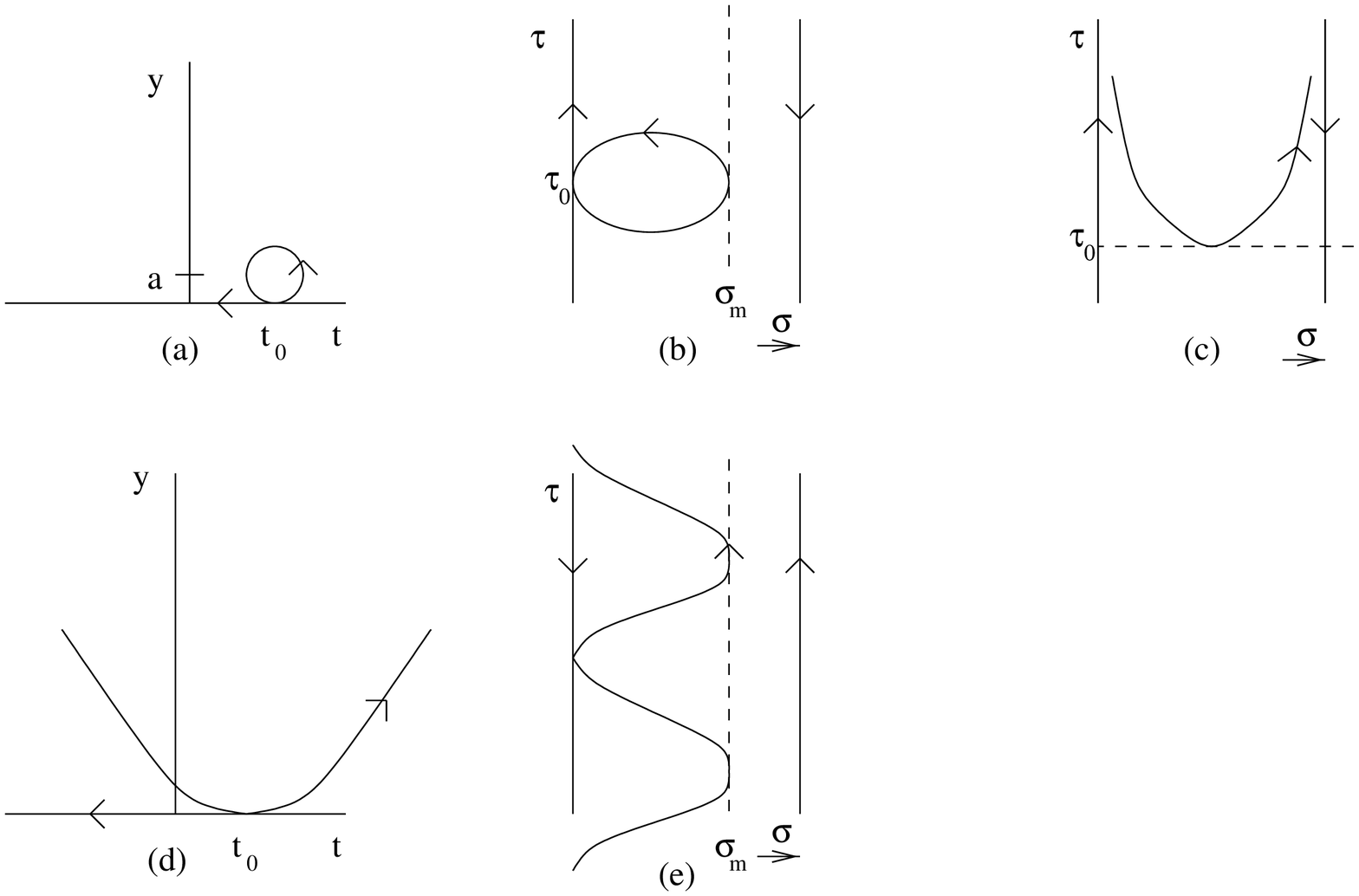}}  
\bigskip
\centerline{\vbox{\noindent {\bf Fig.\ 4.} Trajectories of BPS 
 charged particles
in $AdS_2$ with Euclidean and Lorentzian signatures in various coordinate
systems. The orientation of the arrows indicate the 
charges of the particles and the charges of the boundaries.
 (a) Euclidean trajectory in Poincare coordinates. (b) Euclidean trajectory
in the strip coordinates. (c) Zero energy Euclidean trajectory in strip
coordinates---this will be relevant for tunneling. (d) Lorentzian trajectory
in Poincare coordinates. (e) Lorentzian trajectory in (global) strip coordinates. 
}}
\vskip .8cm

\subsec{Multi-Black Hole Configurations}

The near-horizon two-black hole geometry \ssdt\ can be generalized to
$n$ black holes simply by replacing $V$ in \ssdt\ with 
\eqn\hla{V=\sum_{i=1}^n{Q_i \over |\vu -\vu_i|}.}
The resulting geometry has $n$ \adst\ regions near the $n$ horizons  
$\vx=\vx_i$ as well as an asymptotic one at large $\vx$. 
Timelike singularities lie behind each  of the $n$ horizons.
The 3$n$-dimensional effective action deduced from \fe\ is 
\eqn\rls{S = {3 \over 16 \pi} \int dt \sum_{i,j,k,l} Q_i Q_j
Q_k Q_l |\del_t \vu_i - \del_t \vu_j|^2 \int d^3 U
{(\vu-\vu_i)\cdot(\vu-\vu_j) \over |\vu - \vu_i|^3 |\vu - \vu_j|^3 |\vu -
\vu_k| |\vu - \vu_l|}.}

Black holes encountered in string theory tend to 
involve more than just a single type of charge. This leads to  
non-trivial generalizations of \rls. In five dimensions the 
black holes considered in \ascv\ have three charges $Q_1$, 
$Q_5$ and $n$. The near-horizon geometry of a single black hole 
is \ad $\times S^3$. The moduli space geometry was studied in 
\dkjm. The near-horizon effective action for $p$ black holes is  
\eqn\effact{\eqalign{S = & 
{1\over 4} \int dt 
\sum_{i\neq j}^p \sum_{k}^p 
(Q_{1i} Q_{5j} n_k + Q_{1i} n_j Q_{5k}
+ Q_{5i} n_j Q_{1k}) |\p_t\vu_i-\p_t\vu_j|^2 
\cr & 
\;\; \times
\left[{1\over |\vu_i-\vu_j|^2
|\vu_i-\vu_k|^2} + {1\over |\vu_i-\vu_j|^2 |\vu_j-\vu_k|^2} - {1\over
|\vu_i-\vu_k|^2 |\vu_j-\vu_k|^2} \right] 
,}}
where $U = r\sqrt{R V \over g^2 {\alpha'}^4 }$.  
In the sum, there are divergent terms when $k=i$ or $j$, but they cancel.
Specializing to the case of two black holes, we have
\eqn\gth{\eqalign{S = 
{1\over 2}  \int dt &
\; 
{\Gamma_3 \over |\vu_1-\vu_2|^4} 
|\p_t \vu_1-\p_t \vu_2|^2, }}
where
\eqn\gthdefs{\eqalign{
\Gamma_3 &= Q_{11} Q_{52} n_2 + Q_{51} n_2 Q_{12} + n_1 Q_{12} Q_{52} 
+ Q_{12} Q_{51} n_1 + Q_{52} n_1 Q_{11} + n_2 Q_{11} Q_{51}.}}
The four-dimensional multi-charge expression was found in \jmfd.
Note that, like the single-charge four dimensional case \erf, 
the moduli space is locally flat.  In fact, the two black hole case \gth\ 
is exactly $\IR^4$; $U=\infty$ is an ordinary point,
and the 3-sphere at $U=0$ is asymptotic infinity.

\subsec{The Dual CFT Picture}

Up to this point our discussion has been largely in the framework of 
semiclassical quantum gravity, and has not significantly 
involved string theory. 
In this section we discuss the \ad\ trees 
in terms of the dual CFT on the boundary.

In the case of $AdS_5$ analogous  multi-center gravity solutions 
(and D3 branes in $AdS$) 
correspond  to Coulomb branches in the field theory where the scalar fields
in the vector multiplets
have  expectation values and the gauge symmetry is broken to subgroups. 
Fields in a 0+1--dimensional quantum mechanical system
do not  have well defined expectation values; they fluctuate. 
The $AdS_2$ trees correspond to different classical vacua of a quantum
mechanical theory and we expect that the system moves continuously 
among  them. 

It seems natural to ask whether these trees correspond in any sense to the Coulomb 
or Higgs branches of the theory. In the classical $ g \to  0$ limit of
a D-brane system these two 
branches are very different.
Quantum mechanically the system fluctuates and explores the
whole moduli space.
It has been argued  in similar contexts in \zerotwo \higgsdecoupling\
that the two branches decouple in the low energy limit and that 
the system, if initially in the Higgs branch, explores only 
the Higgs branch and does not wander on to the Coulomb branch. 
The tunneling processes discussed below seem to suggest that the 
branches are not decoupled. 
However this is not the end of the story, since it could be that
branes in $AdS_2$ could correspond to something analogous
to ``small'' instantons or similar configurations in the 
Higgs branch. For example 
in the $AdS_5$ description of Yang Mills field theory on $S^4$
a Yang Mills instanton --- which corresponds to the
Higgs branch of the D(-1) brane gauge theory ---
 is a D(-1) brane moving in $AdS_5$ \instantons .
This implies that D(-1) branes in the near-horizon geometry are already
dissolved in the D3-brane field theory and that we should
interpret their positions as sizes. This analogy is imperfect 
because there is no Coulomb branch at all for the $S^4$ Yang-Mills
theory. Nevertheless it suggests the possibility that 
a brane in $AdS_2$ or an \ad\ tree could be 
corners of the Higgs branch near 
the point where the Coulomb branches meet. This view is corroborated by 
the observation that the volume of the moduli space for a 0-brane 
--- as discussed above ---
in $AdS_2$ is finite near $U_{12}=\infty$.

\newsec{$AdS$ Fragmentation}

In the previous section we discussed configurations in which a single 
charge $Q_1+Q_2$ \adst\ can branch into two \adst\ spaces 
with charges $Q_1$ and $Q_2$  as one moves spatially from 
infinity towards the horizon. As can be seen from the analysis of 
geodesics in section 3.3, this branching point can actually reach the
boundary of \adst\ in finite global time. In principle the 
geometry could then fragment into two completely separate 
\adst\ universes with charges $Q_1$ and $Q_2$. Whether or not this actually 
happens depends on the boundary conditions at the \adst\ boundary.

In this section we will consider some examples in which fragmentation 
of a single $AdS$ universe into several smaller $AdS$ 
universe does 
occur, for both \ad\ as well as higher dimensional examples. 
The processes we consider are tunneling processes mediated by 
topology-changing instantons. These or closely related instantons 
were considered in a different context in 
\refs{\teitel,\dggh}. The first two subsections 
consider processes in which an initial $AdS$ space fragments into one 
macroscopic and one  
microscopic component. The microscopic component is described by 
a brane. We then turn to the 
case (analyzed by Brill \brl ) 
in which it splits into two macroscopic $AdS$ universes. 

\subsec{The Non-Supersymmetric Case}

Consider $AdS_D$ for general $D$ endowed with 
a constant antisymmetric $D$-form field strength. 
In flat space, a constant antisymmetric $D$-form field strength
leads to $(D-2)$-brane creation for any variety of  
$(D-2)$-brane that is charged under the 
$D$-form field strength \dggh . 
In the case of $D=2$ the
field strength is a two-form and this reduces to the  
well known Schwinger pair production of 0-brane anti-0-brane pairs. 
The Schwinger process is described by an instanton in which the 
charged particle moves
in a circular trajectory in the electromagnetic field. 
The vacuum decay rate is proportional to $e^{-S_{e}}$ where
$S_e$ is the action of this Euclidean solution. 
The configuration to which the vacuum decays --- namely a 0-brane 
anti-0-brane pair --- is found 
by cutting the instanton in half at the moment of time symmetry (say $\tau=0$).  
The branes subsequently accelerate off to infinity. 
For general $D$ the analogous Euclidean solution is a $D-1$ sphere. 
Cutting it in half we get at $\tau=0$ a $D-2$ sphere. In the subsequent Lorentzian
evolution the sphere expands due to the force
exerted by the $D$-form field strength. In flat space this process
of brane creation screens the $D$-form field strength. 
In $AdS_D$ ($D-2$)-branes can also be created in this fashion \teitel .

We will here describe these instantons in the test-brane approximation,
where we neglect the charge of the brane compared with the total flux
of $F_D$ in $AdS_D$.  
Let us write the metric of $AdS_D$ as
\eqn\metric{
ds^2 =  R^2 \left(\cosh^2\!\rho \, d\tau^2 + d\rho^2 + \sinh^2\!{\rho} \, d\Omega_{D-2}^2 
\right),}
where $R$ is the anti-de Sitter radius.
Then the action of a spherically symmetric 
 brane coupled to the $D$-form field strength
is 
\eqn\act{
S = TR^{D-1} \Omega_{D-2}  \int d\tau \left[
\sinh^{D-2}\rho  \sqrt{ \cosh^2\rho + \left({d \rho
 \over d \tau}\right)^2 } - q 
\sinh^{D-1}\rho  \right]
,}
where $T$ is the brane tension,
 $q$ is the ratio of the charge of the brane to its tension
and 
\eqn\dvm{\Omega_{D-2}={2 \pi^{{(D-1)\over 2}} \over \Gamma({D-1 \over 2})} }
is the volume of a unit $D-2$ sphere. 
In the BPS case the forces balance so $q=1$. In a supersymmetric 
theory the BPS bound 
implies that $ q \leq 1$ for all possible branes. In a non-supersymmetric 
context $q >1$ is possible (for example the electron). In the next subsection 
an example is given in string theory. 

Now we turn to solutions of the brane action \act\ for $q>1$. 
Since the action \act\ is independent of time, Euclidean energy is conserved.  
For a spherically symmetric and compact surface, this energy is
zero.
Energy conservation
then implies 
\eqn\geneq{
{ \cosh^2 \rho \over \sqrt{\cosh^2\rho + \dot \rho^2 }} - q \sinh\rho = 0
}
which is independent of $D$.
The solution of this equation is 
\eqn\inst{
 \cosh\rho = { \cosh \rho_{max}\over \cosh \tau },}
where $\tanh \rho_{max} = 1/q$. Equation \inst\ describes a closed 
$D-1$ surface with maximum radius $\rho_{max}$. 
The action of this instanton is 
\eqn\sinst{\eqalign{ 
S_{inst} &=  { 2  T R^{D-1}
 \Omega_{D-2}\over \sinh \rho_{max}} \int_0^{\rho_{max}}
d \rho { \sinh^{D-2} \rho \sqrt{ \sinh^2 \rho_{max} - \sinh^2 \rho } \over
\cosh \rho }\cr&={ \pi^{D\over 2} T R^{D-1} \over \Gamma({D+2 \over 2})}
\sinh^{D-1}\rho_{max}F(1,{D-1 \over 2};{D+2 \over 2};-\sinh^2\rho_{max}).}
}

At $\tau=0$ one can match \inst\ to the Lorentzian solution
\eqn\instlor{
\cosh\rho = { \cosh\rho_{max} \over \cos \tau },
}
which describes the post-tunneling evolution. Note that the brane gets to the 
boundary (at $\rho = \infty$) in 
finite time (at $ \tau= \pi/2 $) (see fig.\ 5). 

\vbox{
\vskip .6cm
\centerline{
\epsfxsize=0.4\hsize\epsfbox{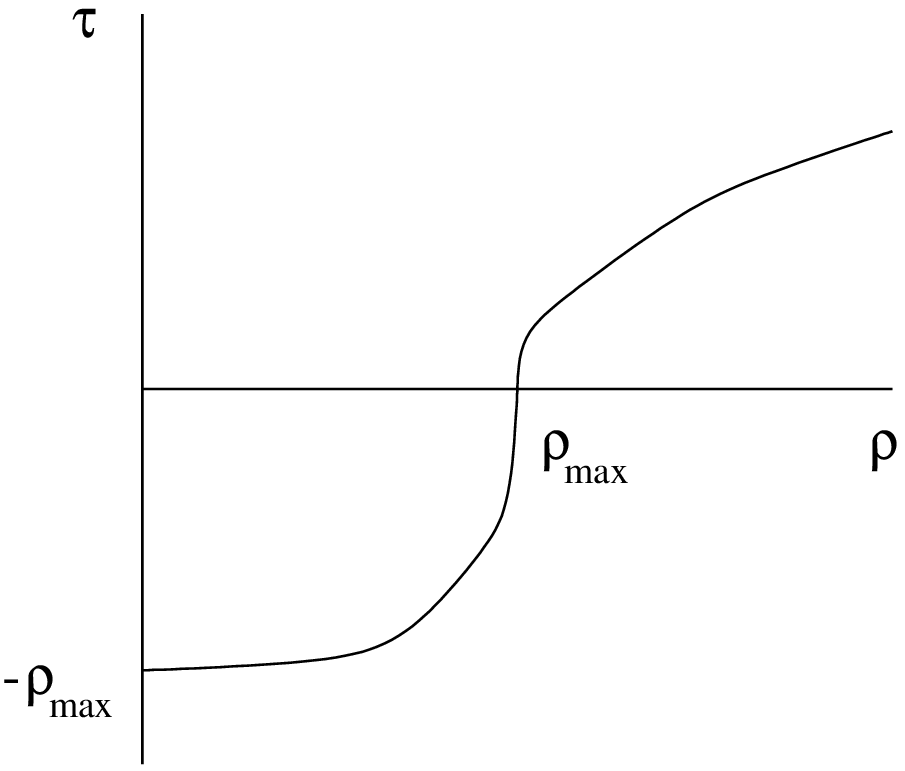}}  
\bigskip
\centerline{\vbox{\noindent {\bf Fig.\ 5.} 
The Euclidean instanton~\inst\ for $\tau<0$ matches onto the Lorentzian
solution~\instlor\ at $\tau=0$.
}}
\vskip .7cm
}
If $q<1$ the Euclidean solution is 
\eqn\qless{
\cosh \rho = { \sinh \tau_{max} \over \sinh \tau }~~~~~~~ \tanh \tau_{max} 
 \equiv  q 
}
It does not describe a tunneling process because the solution 
does not have  a moment of time symmetry and also its action is
infinite. 
This is expected since tunneling is forbidden by energy conservation 
for $q < 1 $.

\subsec{An Example of a Brane With Charge Greater Than Tension}

In this section we will give an example of a non-supersymmetric
 $AdS_3\times S^3\times K3$ compactification 
which is unstable to fragmentation.
Consider type IIB on $K3$. The six dimensional theory has an $SO(5,21)$
multiplet
of strings, coming from branes wrapped on various cycles of $K3$. 
A string is characterized
by a charge vector $q^I$ transforming in the vector  under $SO(5,21)$. 
If $q^2 >0 $ the string is BPS and the near-horizon geometry 
is $AdS_3 \times S^3 \times K3 $. If $q^2 <0$ the
string will not be BPS, but nevertheless there is a supergravity solution and
its near-horizon geometry is again $AdS_3 \times S^3 \times K3 $\dpop.
A simple example is the following. Take a set of $Q_5$ D5-branes wrapped
on $K3$ which leads to a string in six dimensions. We can add D1-branes 
to this system. Supersymmetry is preserved only if the 
D1-branes have the right charge --- we choose it to be positive.
The BPS bound for the tension of the string is  
$T \sim  | Q_5 V_4 + Q_1 |/g $ where $V_4$ is the volume of $K3$
in string units. 
When we wrap a D5-brane on $K3$ there is one unit of negative D1-brane
charge induced on the brane~\vafainduc. So there are BPS strings with
 charges
$(q_5, q_1) = ( 1,-1)$ whose tension is $T_{(1,-1)} = (V_4 -1)/g$ (we
are assuming $V_4>1$). 
Supersymmetric black strings  will have charges $Q_1 \geq 0 $.  
Consider a black string whose charge is $Q_1 < 0$. 
The bosonic part of the supergravity equations have a structure which is
not very sensitive to the relative sign of $Q_1$.
 Actually the 
only change in the solution is the sign in the electric part of the $B$ field. 
In particular, the volume of the $K3$ at the horizon, (encoded in the
 dilaton), will be $V = |Q_1|/Q_5$
so we will take $|Q_1| > Q_5$ so that the above formula for the tension
of the $(1,-1)$ string is correct. 

We can also study small fluctuations around this solutions and check that
there is no tachyon producing an instability (i.e. all tachyons
obey the  Breitenlohner-Freedman bound $  m^2 \geq -1 $). 
The basic reason is that the equations involve 
a coset space $SO(5,21)/SO(5)
\times SO(21)$ and  self-dual field strengths transforming in the {\bf 5}
of $SO(5)$ and antiself-dual field strengths transforming in the {\bf 21}.
In the case that $Q_1 >0$ only the self-dual field strength is non-zero
while if $Q_1<0$ only the antiself-dual part is non-zero in the near
horizon region. The bosonic equations  are symmetric under the 
exchange of self-dual and antiself-dual fields together with the change
of the $SO(5)$ and $SO(21)$ pieces of the coset. 
So if for $Q_1>0$ we had 21 fixed scalars and a similar number 
of ``good'' tachyons now we will have 5 fixed scalars and a similar
number of ``good'' tachyons, etc. 
In the negative $Q_1$ case the string with charges 
$(q_5, q_1) = ( 1,-1)$ will feel a repulsive force, since  the only
 difference 
with the supersymmetric case (in which the forces balance) is that the
 onebrane electric force is repulsive rather than attractive.
So this string will have $q>1$ in the notation of the previous
section (more precisely $q = {|Q_1| + Q_5 \over |Q_1| - Q_5 }$).
 This implies that the black hole would lose its
charge by emitting these $q>1$ branes. 

It is natural to ask whether other non-supersymmetric $AdS$ spaces
would have similar instabilities. These kind of instabilities 
would generically occur for $AdS_2$ cases; a well known  example is
an extremal electrically charged black hole in our universe, which
discharges by emitting electrons (which have $q \sim  2 \times  10^{21}$). 
It can be seen that if we have an $AdS_5\times M^5$ compactification 
where $M^5$ is an Einstein manifold, then the gravity equation
of motion implies  that a three-brane moving in that geometry will have
$q=1$ regardless of whether supersymmetry is broken or not. 
There could be, however, other branes with $q>1$ if the Einstein 
manifold has some small cycles on which one can wrap branes etc. 
In cases where one has a ``warped'' geometry \warp , i.e. a solution where
the radius of $AdS_5$ depends on the coordinates on $M^5$ then
the threebrane could have $q>1$. An example is the
compactification that has $SO(5)$ symmetry that arises as an unstable
IR ``fixed point'' after perturbing the ${\cal N} =4 $ theory by
a relevant operator \sofive .  

The nature of the tunneling process is illuminated by consideration 
of the energy of a (momentarily) static spherical brane at radius $\rho$.  
This is not 
an equilibrium configuration. 
This energy has a positive contribution from the
mass of the brane and a negative contribution from
the electric potential:
\eqn\energy{
E(\rho) = T R^{D-2}\Omega_{D-2} \left[ \sinh^{D-2} \rho \cosh \rho - q \sinh^{D-1}
\rho \right]
}
We see that if $q>1$ then $E(\rho)\to -\infty $ as $\rho \to \infty$.
So the system decays in order to reach this lower energy configuration. 
If $q=1$ then
we see that $E(\rho) \to \{ \infty, const, 0\}$  for the 
$\{ D>3, D=3,D=2 \}$ cases as $\rho \to \infty$. 
This is consistent with the fact that only for $D=2$ can we
have tunneling in the $q=1$ case. 

It is interesting to note that in the $AdS_3$ case the constant
value of the energy is related to the change
in the central charge of the system when we remove a brane.
This is anticipated from the dual NS-sector CFT description 
in which the ground state energy is proportional to the 
central charge.  
More precisely, if we have a D5-D1 brane system then
the change in energy is $Q_1/2$ if we remove a fivebrane
and $Q_5/2$ if we remove a onebrane.
This further implies that if we have enough energy above the ground state
in $AdS_3$ then the system can decay by emitting branes. 
In the case with NS charges (NS fivebranes and fundamental strings) 
this decay mode might be related  to the negative norm states 
for the $SL(2,\IR)_{Q_5} $ WZW model arising 
when the  square of the  mass of the state is of the order of $Q_5$.

\subsec{The Supersymmetric Case}  

In this subsection we consider the supersymmetric case $q=1$.
The solution to the geodesic equation for $q=1$  is 
\eqn\qone{
e^\tau  = \cosh \rho
}
as discussed previously in \zeroenstrip . 
Unlike the $q<1$ case \qone\ is not a ``bounce'' solution. 
It does not have a moment of time symmetry and there is no negative mode 
indicating an instability. 
Rather it represents tunneling between two degenerate vacua. 
At $\tau=-\infty$ the instanton is asymptotic to charge $(Q_1+Q_2)$ \ad, 
while at $\tau=+\infty$ one has charge $Q_2$ \ad\ plus a charge $Q_1$ 
brane in the boundary. 

For $D>2$ the instanton action is infinite so the tunneling does not actually occur.
For $D=2$ it takes the finite value
\eqn\fvl{S_{instanton}= \pi Q_1Q_2.}
Comparing with \etl\ we see that this can be written 
\eqn\oty{S_{instanton}=-{1 \over 2}  \Delta S_{BH},}
where $\Delta S_{BH}$ is the difference in the Bekenstein-Hawking 
entropy of the initial and final 
states. 
In a description which microscopically accounts for the Bekenstein-Hawking 
entropy, twice the factor \oty\ would arise in transition 
$probabilities$ from averaging over initial and summing 
over final states. This agrees with the fact that the instanton
\oty\ gives transition $amplitudes$. Apparently the instanton mysteriously 
knows the number of microstates. Previous examples of instantons counting 
microstates in this fashion can be found in \ggs, \brl.

All supersymmetric instantons will have fermion zero modes which we have 
not analyzed in detail. This means that the transitions will be
accompanied by a change in fermion number and/or spacetime momentum,%
\foot{This momentum refers to the original asymptotically flat region. In 
the $AdS$ context it is conjugate to zero modes of singleton fields.}
and will not shift the ground state energy.

\subsec{The Brill Instanton}

A tunneling process in which an initial \adst\ universe 
fragments into two final \adst\ universes 
 should be described by a smooth 
instanton with one initial and two final \adst\ boundaries. 
Such an instanton was discovered by Brill \brl, whose work we review in 
this subsection. In the limit in which one of the final universes is small
and can be treated like a brane, this instanton reduces to
\qone\
of the previous subsection. 

The Euclidean action for the Einstein-Maxwell theory is 
\eqn\sfg{S=-{1 \over 16 \pi}\int d^4x\sqrt{g}\bigl(R-F^2\bigr)
-{1 \over 8 \pi}\oint d^3x \sqrt{h} K,}
where $K$ and $h$ are the trace of the extrinsic curvature and 
the induced metric on the boundary and we have set Newton's constant to 
one. This has the family of solutions
\eqn\ssdk{\eqalign{ds^2&=V^2d\vx^2+V^{-2}dw^2, \cr
                   *F&=- dw\wedge dV^{-1},\cr
                    \vec{\nabla}^2 V(\vx)&=0,}}
where $\vec{\nabla}^2$ is the Laplacian on flat $\IR^3$.  The special case 
\eqn\rbu{V={Q_0 \over |\vx|}}
corresponds to the Euclidean \adst\ \rb\ universe with magnetic charge 
$Q_0$ on the \st, and \ad\ cosmological constant $2 \over Q_0$. 
The Brill instanton is\foot{ This metric is obtained by 
analytic continuation $t\to iw$ of \ssdt, but we will not 
interpret $w$ as Euclidean time.}
\eqn\bint{V={Q_1 \over |\vx -\vx_1|}+{Q_2 \over |\vx -\vx_2|}.}
For $\vx \to (\vx_1, \vx_2, \infty)$ the metric given by \bint\ 
approaches the \adst\ metric \rbu\ with charge $Q=(Q_1,Q_2,Q_0=Q_1+Q_2)$.

We wish to interpret this as a semiclassical contribution to the 
tunneling of an initial charge $Q_0=Q_1+Q_2$ \adst\ spacetime to final 
charge $Q_1$ and $Q_2$ spacetimes. In order to 
do so we need to identify one initial surface $\Sigma_0$ 
and two final surfaces $\Sigma_1$ and $\Sigma_2$ 
with topologies $\IR\times S^2$ corresponding to spatial slices
of \adst\ spacetimes with the appropriate charges. 
The metrics on the surfaces should agree with those of the
corresponding slices of \adst. The extrinsic curvatures should vanish 
so that the continuation back to Lorentzian signature gives real initial
data.\foot{The surfaces given in \brl\ differ from those described here and do 
not satisfy this criterion.} 

\vskip .6cm
\centerline{
\epsfxsize=0.25\hsize\epsfbox{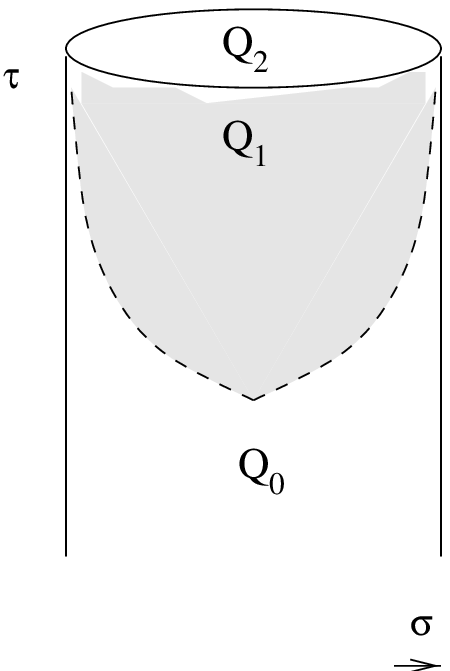}}  
\bigskip
\centerline{\vbox{\noindent {\bf Fig.\ 6.} 
The Brill instanton. We start with an charge $Q_1 + Q_2$ $AdS_2 \times S^2$ space
which splits into two $AdS_2 \times S^2$ spaces of charges  $Q_1$ and $Q_2$. 
For finite Euclidean time they still look like a single charge
$AdS$ space close to the boundary. Only at infinite time do the 
two $AdS$ spaces separate. 
}}
\vskip .7cm

Let us introduce the following variable
\eqn\genz{
y = \left( { Q_1 \over \sqrt{|\vec x - \vec x_1| } } + { Q_2 \over
\sqrt{|\vec x - \vec x_2 | }}  \right)^2
}
And then define the variables $\tau,\sigma$ through 
\eqn\newvar{
 w + i y = e^{\tau + i \sigma }
}
Where $ 0 \leq  \sigma \leq \pi $. We see that for
$\tau \to - \infty $ we have the single charge $Q_0$ $AdS$ space
in the (Euclidean) strip coordinates  and
for large $\tau $ we have two $AdS_2$ spaces of charges 
$Q_1$ and $Q_2$  which actually meet at $\sigma \to 0,\pi$ to form again
a charge $Q_0$ $AdS_2$ space (the change of variables \genz\ is, in a sense, 
double valued). The point $\sigma$ where they meet 
goes to the boundary ($\sigma = 0, \pi$) when
$\tau \to + \infty $ so that in the limit we really have two disconnected
$AdS_2$ spaces. 
We need to regulate the spatial extent of $AdS_2$ (the $\sigma$ coordinate)
and also the temporal extent. This could be achieved 
by taking cutoffs $ \epsilon < \sigma < \pi - \epsilon $ 
and $-T < \tau < T$. Notice that if we take first a finite $\epsilon$,
then the two final $AdS_2$ regions become disconnected at finite $T$. 

The tunneling amplitude is proportional to the exponential 
of minus the instanton action.  Reducing the action to a surface term 
and subtracting the 
action of the vacuum to vacuum instanton, Brill finds   
an amplitude proportional
\eqn\trdg{A_{Q_0\to Q_1+Q_2}\sim e^{\half\Delta S_{BH} },}
where
\eqn\trg{\half \Delta S_{BH}=\half \{S_{BH}(Q_1) + 
S_{BH}(Q_2)-S_{BH}(Q_1+Q_2)\}=-\pi Q_1Q_2.}
In this expression 
\eqn\sbh{S_{BH}(Q) =\pi Q^2}
is the Bekenstein-Hawking entropy for a charge $Q$ extremal black hole. 
Squaring the amplitude to get the transition probability one finds that 
it is proportional to minus the exponential of the entropy decrease, as 
expected. This result agrees exactly with the result \fvl\ computed 
for $Q_1 \ll Q_2$ . Because of the necessity of subtractions, it is 
not manifestly obvious (although it is expected) 
that the action will be the same when computed 
for the initial and final surfaces described above.

An interpretation of the tunneling process which does not refer to the 
decoupled asymptotically flat region can  
be given in the context of 
third-quantized Hilbert space in which states are labeled 
by the occupation numbers 
$n_i$ of \adst\ spacetimes with charges $Q_i$. 
The Brill instanton 
corresponds to a nonperturbative correction to the Hamiltonian which 
changes these occupation numbers. Due to charge conservation there are 
superselection sectors labeled by the total charge. In the 
semiclassical approximation 
considered here, the \adst\ spacetimes are like non-relativistic
particles. 
There is 
no pair-creation of oppositely charged spacetimes, and $Q_i$ is 
restricted to be positive.

In general 
topology-changing processes which change the 
number of universes are problematic (a review can be found in \astasi). 
Among other reasons, 
it is difficult to describe such processes by a Hamiltonian 
because there is in general no canonical way to compare the times 
of different universes.  This  
problem cannot arise in 
the present context because it was derived as a limit of a system which
included the asymptotically flat region and  
did have a Hamiltonian. The separate \adst\ universes carry a preferred 
time with them as a remnant of the asymptotically flat region 
which once joined them.

\centerline{\bf Acknowledgements} 

  We have benefitted from useful conversations with 
R.\ Britto-Pacumio, J.\ de Boer, M.\ Headrick, K.\ Hori, 
H.\ Ooguri, E.\ Silverstein, M.\ Spradlin, J.\ Stopple, L.\ Susskind and
E.\ Witten.
This work was supported in 
part by an NSF Graduate Fellowship, an NSERC PGS B Scholarship
and DOE grant DE-FGO2-91ER40654.
\listrefs
\end